\documentclass[10pt]{article}

\usepackage[]{rw_emulateapj}
\usepackage[]{apjfonts}

\begin{document}

\submitted{Accepted for publication in {\it The Astrophysical Journal}}

\title{Identification of a nearby stellar association in the
Hipparcos catalog: implications for recent, local star formation.}

\author{B. Zuckerman\altaffilmark{1} \& R. A. Webb\altaffilmark{1}}
\affil{Dept. of Astronomy and Physics - Univ. of California, Los
Angeles  \\ 8371 Math Science Bldg. - Box 951562 \\ Los Angeles,
CA 90095-1562 \\ ben@astro.ucla.edu; webbr@astro.ucla.edu}

\altaffiltext{1}{Visiting astronomer, Cerro Tololo Inter-American
Observatory (CTIO) which is operated by AURA Inc. under contract to the
National Science Foundation}

\begin{abstract}

The TW Hydrae Association ($\sim$55 pc from Earth) is the nearest
known region of recent star formation.  Based primarily on the
Hipparcos catalog, we have now identified a group of 9 or 10 co-moving
star systems at a common distance ($\sim$45 pc) from Earth that appear
to comprise another, somewhat older, association (``the Tucanae
Association'').  Together with ages and motions recently determined
for some nearby field stars, the existence of the Tucanae and TW
Hydrae Associations suggests that the Sun is now close to a region
that was the site of substantial star formation only 10-40 million
years ago.  The TW Hydrae Association represents a final chapter in
the local star formation history.

\end{abstract}


\section{Introduction}

        For many years, the only known star clusters within 70 pc of
Earth were the rich Hyades (at $\sim$ 45 pc) and the sparce U Ma
nucleus (at $\sim$25 pc).  Then a group of stars $\sim$55 pc from
Earth was established as a {\it bona fide} association of T Tauri
stars of age $\sim$10 million years (Webb et al 1999; Sterzik et al
1999; Webb, Reid, \& Zuckerman 2000).  This TW Hydrae Association
(hereafter TWA) was unrecognized for many decades in spite of its
being the nearest region of recent star formation (Kastner et al
1997). A ubiquitous signpost of newly formed stars has been a nearby
molecular cloud.  But no interstellar cloud has been found near TW Hya
despite multiple searches. Therefore, we ask: do additional unrealized
young associations far from molecular clouds exist near Earth?

        Different teams have used the Hipparcos catalog to search for
previously unrecognized stellar associations. For example, Platais et
al (1998) undertook ``A search for star clusters from the Hipparcos
Data'' and list basic data for 5 ``very likely'' new clusters and
associations as well as 15 ``possible'' ones. Yet, in their own words,
``At distances $<$100 pc the survey is incomplete as a result of the
chosen search strategy.'' Of the 20 potential new groupings in their
Table 1, the closest, which contains 11 members from Hipparcos, is 132
pc away.  

    By contrast, the ``Tucanae Association'' we propose in the present
paper is only $\sim$45 pc from Earth and much younger than the U Ma
and Hyades clusters.  Just as the sparse U Ma cluster nucleus is
accompanied by more numerous U Ma stream stars, we suggest that the
Sun is embedded in a stream of stars with similar space motions (the
``Tucanae Stream''), with the Tucanae Association playing a role
analogous to the U Ma nucleus.  Figures 1a and 1b depict the Tucanae
Association and some stream stars, respectively.  These stars likely
represent some of the younger, nearer, members of the more extensive
Pleiades group or Local Association proposed by Eggen (e.g., see
Jeffries 1995 and references therein).

\begin{figure*}
\figurenum{1}
\epsscale{0.5}
\plotfiddle{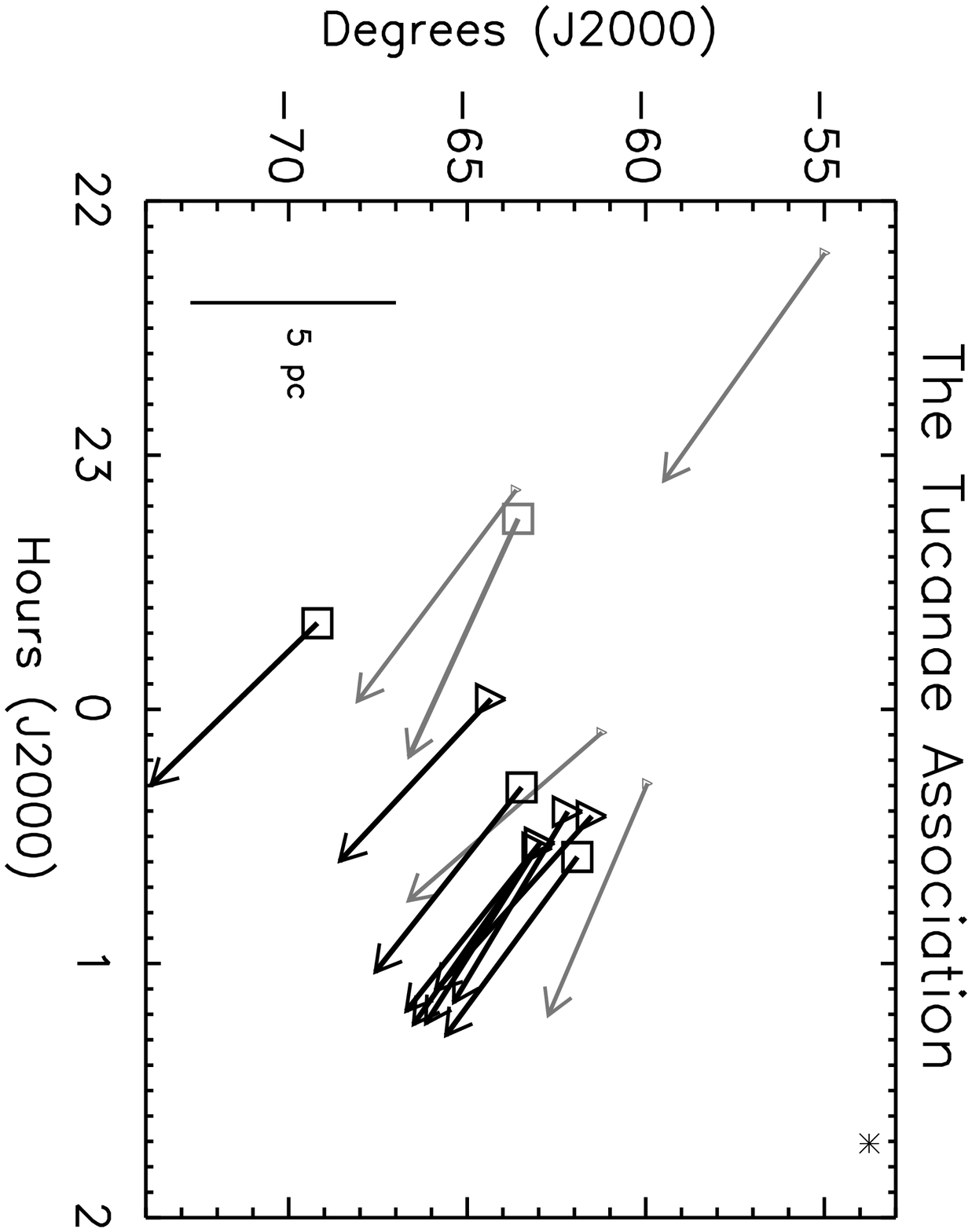}{1.5in}{90}{25}{25}{-70}{-28}
\plotfiddle{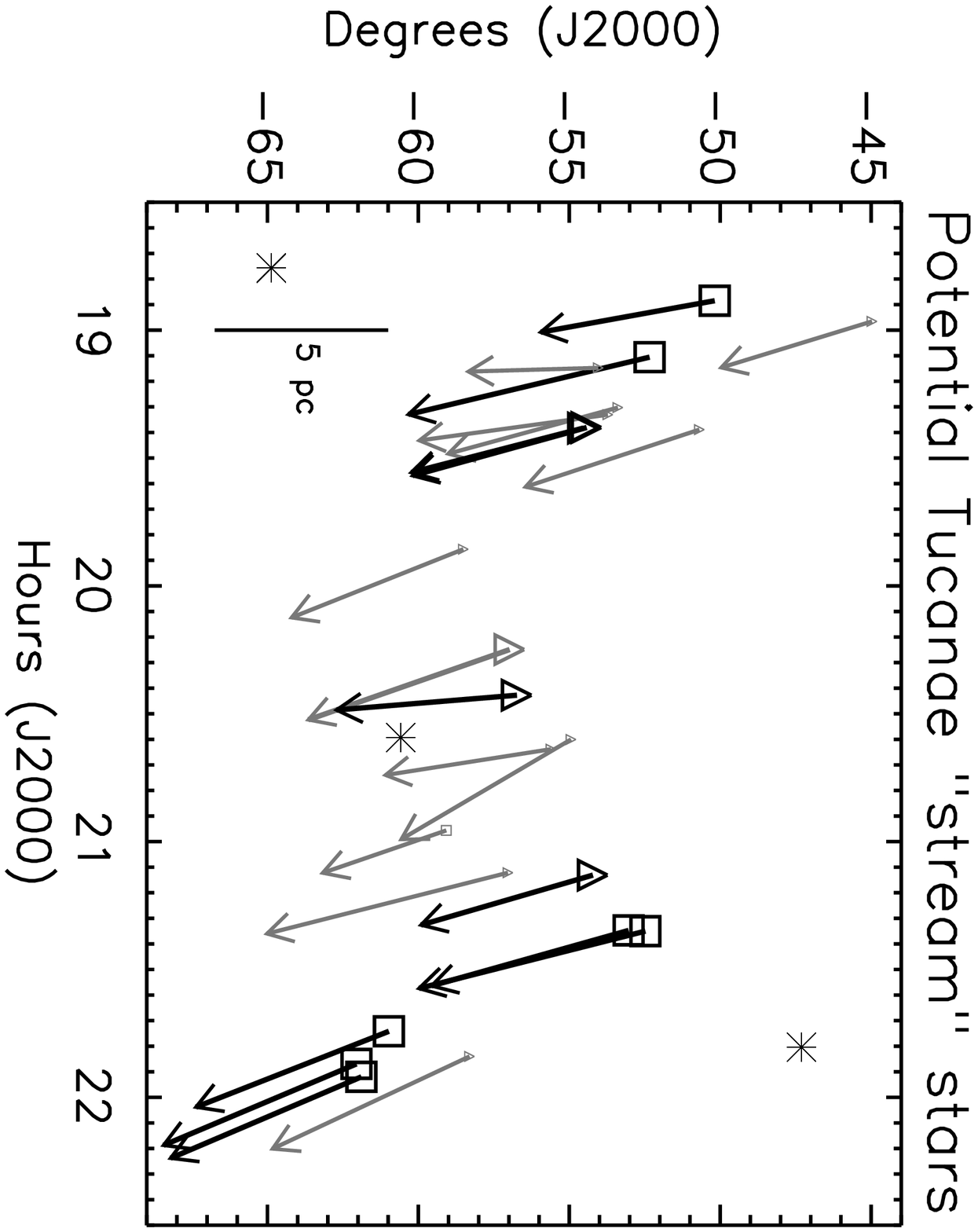}{0in}{90}{25}{25}{95}{-7}
\plotfiddle{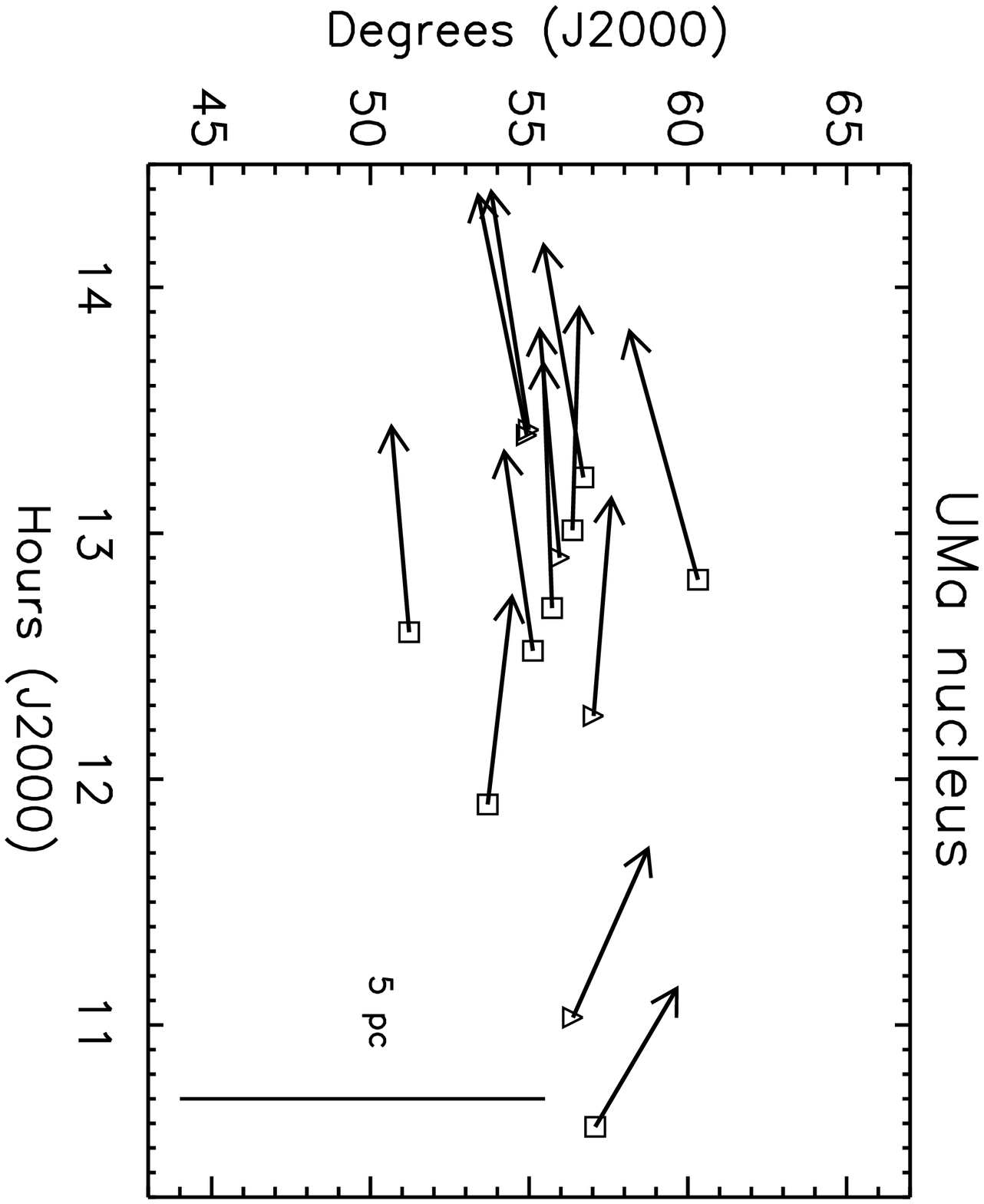}{0in}{90}{25}{25}{260}{14}

\caption{Potential moving group members in Table 1 (panels a. -
The Tucanae Association or nucleus and b. - some of the Tucanae stream
stars) and U Ma nucleus stars (panel c.) shown for comparison.  The
R.A. and Dec axes are scaled to roughly show the true appearance on
the sky though a distortion exists due to the square grid used.  The
vectors represent the proper motion over 250,000 years. Probable
members of the Tucanae Association and stream are indicated with large
symbols and dark vectors; improbable members with small symbols and
light vectors.  Two ``possible'' members are shown with large symbols
and light vectors.  The bar represents 5 pc at the approximate
distance to each group.  X-ray (RASS-BSC) sources are marked with
squares, non-ray sources with triangles. Positions of four additional
young stars with similar space motions which appear in the foreground
of the Association are indicated by asterisks (see text).}
\end{figure*}

\bigskip

\section{Observations}

         Excepting TW Hya itself, the first T Tauri stars in the TWA
were identified in a study of IRAS sources at high galactic latitude
(de la Reza et al 1989; Gregorio-Hetem et al 1992).  It has been shown
that, on the main sequence, young stars are more likely 60 $\mu$m IRAS
sources than are old stars (e.g., Jura et al 1993 \& 1998; Silverstone
et al 2000) and young stars are also more apt to be members of
associations than are older stars. Therefore, we interrogated the
Hipparcos catalog within a six degree radius of two dozen stars
detected by IRAS at 60 $\mu$m and scattered around the sky.  The stars
were those with, in our opinion, reliable excesses at 60 $\mu$m listed
by Mannings \& Barlow (1998), Backman \& Paresce (1993), and/or
Silverstone et al. (2000).  We searched for Hipparcos stars with
similar proper motions and distances to the IRAS stars; results for
our most interesting regions are presented in Figure 1, Tables 1 \& 2
and are discussed below.  Results at additional IRAS stars will be
discussed in a later paper.

    To verify or deny common space motions of Table 1 stars with
similar distances and proper motions in the Hipparcos catalog, we
measured radial velocities with the Bench Mounted Echelle (BME)
spectrograph on the 1.5-m telescope at the Cerro Tololo Interamerican
Observatory (CTIO).  The spectra cover from $\sim$ 5000 - 8200\AA~at a
typical measured resolution of 0.15\AA.  The data, obtained 21-26 (UT)
August 1999, were reduced and calibrated in IRAF.

    Six radial velocity standard stars were observed during the run
with spectral types ranging from F6 to M1.5.  Radial velocity was
determined by cross-correlating the spectrum of the target and a
standard star of similar spectral type observed close in time to the
target.  Approximately 15 echelle orders, chosen to produce strong
correlations and have few atmospheric features, were used to compute
the correlation.  The accuracy of these measurements is strongly
dependent on the S/N of the spectra and the spectral type and
rotational velocity of the target star.  The uncertainty ranges from
$\sim$ 1-10 km/s, the majority being $<$2 km/s.  H$\alpha$ line
profiles and equivalent widths of lithium (6708\AA) were also measured
and are listed in Table 2.

    Projected rotation velocities ({\it v sin i}) were measured and are
listed in Table 2.  We used a procedure similar to that of Strassmeier
et al (1990), specifically we followed the prescription described in
their Section IIId.  We measured the FWHM of lines in the 6420\AA
~region and corrected them for the instrumental broadening (0.134
km/s, FWHM) determined from five ThAr lamp lines at similar
wavelengths.  Following Strassmeier et al, we assumed a macroturbulent
velocity, zeta, equal to 3 km/s.  For a few of the faint late K-type
and M-type stars, {\it v sin i} is quite uncertain due to low S/N.

    With the 0.9-m telescope at CTIO we obtained BVRI photometry of
stars in Table 1.  The purpose was to determine whether any of the
stars were sufficiently young to still be above the main sequence.
These data will be reported in a later paper.

\begin{deluxetable}{llllccccccccc}

\tablefontsize{\scriptsize}

\tablenum{1}
\tablecolumns{13}  

\tablecaption{Potential members of the Tucanae Association \label{Tab1}}
\tablehead{
    \multicolumn{3}{c}{Catalog number} &
    \colhead{Nuclear}    &
    \multicolumn{2}{c}{J2000.0}  &
    \colhead{}    &
    \colhead{}  &
    \colhead{}    &
    \colhead{ROSAT}  &
    \colhead{$\pi$}    &
    \multicolumn{2}{c}{Prop Mot (mas yr$^{-1}$)} 
\\
    \colhead{HIP}   &
    \colhead{HD}   &
    \colhead{HR}     &
    \colhead{member?}     &
    \colhead{R.A.}    &
    \colhead{Dec}    &
    \colhead{m$_v$}    &
    \colhead{B-V}  &
    \colhead{SpT}    &
    \colhead{cps}    &  
    \colhead{mas}    &  
    \colhead{$\alpha$}    &  
    \colhead{$\delta$}    
}
    
\startdata

\cutinhead{1a.  Probable and possible members of the Tucanae nucleus
or stream}

1481   & 1466    & \nodata & $\surd$ & 00 18 26.0 & -63 28 38
& 7.5  & 0.54    &
F8     & 0.26    & 24.42   & 90  & -56  
\nl

1910   & \nodata & \nodata & $\surd$ & 00 24 08.9 & -62 11 04
& 11.3 & 1.39    &
M0     & \nodata & 21.59   & 93  & -46  
\nl

1993   & \nodata & \nodata & $\surd$ & 00 25 14.6 & -61 30 48
& 11.3 & 1.35    &
K7     & \nodata & 26.69   & 85  & -63  
\nl

2484   & 2884    & 126     & $\surd$ & 00 31 32.6 & -62 57 29
& 4.37 & -0.07   &
B9     & \nodata & 23.35   & 92  & -59  
\nl

2487   & 2885    & 127     & $\surd$ & 00 31 33.4 & -62 57 56
& 4.54 & 0.15?   &
A2+A7& \nodata & 18.95   & 98  & -51  
\nl

2578\tablenotemark{a}   
       & 3003    & 136     & $\surd$ & 00 32 43.8 & -63 01 53
& 5.09 & 0.04    &
A0     & \nodata & 21.52   & 94  & -51  
\nl

2729   & 3221    & \nodata & $\surd$ & 00 34 51.1 & -61 54 58
& 9.6  & 1.05    &
K4     & 0.5     & 21.78   & 87  & -59  
\nl

92680  & 174429  & \nodata & \nodata & 18 53 05.9 & -50 10 49
& 7.8(8.5) & 0.78&
K0     & 1.0    & 20.14   & 17  & -85 
\nl

93815  & 177171  & 7213    & \nodata & 19 06 19.9 & -52 20 26
& 5.2  & 0.53    &
F7     & 2.08    & 19.07   & 26  & -117 
\nl

95261\tablenotemark{a}  
       & 181296  & 7329    & \nodata & 19 22 51.2 & -54 25 25
& 5.05 & 0.02    &
A0     & \nodata & 20.98   & 28  & -83  
\nl

95270\tablenotemark{a}  
       & 181327  & \nodata & \nodata & 19 22 58.9 & -54 32 16
& 7.0  & 0.48    &
F5     & \nodata & 19.77   & 28  & -82  
\nl

99803\tablenotemark{b}  
       & 191869  & \nodata & \nodata & 20 14 56.1 & -56 58 34
& 7.24 & 0.49    &
F6.5   & \nodata & 15.32   & 33  & -92  
\nl

100751 & 193924  & 7790    & \nodata & 20 25 38.8 & -56 44 06
& 1.94 & -0.12   &
B7     & \nodata & 17.60   & 7   & -87  
\nl

104308 & 200798  & \nodata & \nodata & 21 07 51.2 & -54 12 59
& 6.7  & 0.24    &
A5     & \nodata & 15.05   & 28  & -82  
\nl

105388 & 202917  & \nodata & \nodata & 21 20 49.9 & -53 02 02
& 8.6  & 0.69    &
G5     & 0.62    & 21.81   & 30  & -94  
\nl

105404 & 202947  & \nodata & \nodata & 21 20 59.8 & -52 28 39
& 8.9  & 0.85    &
K0     & 0.67    & 21.72   & 34  & -105 
\nl

107345 & \nodata & \nodata & \nodata & 21 44 30.1 & -60 58 38
& 11.7 & 1.4     &
M1     & 0.137   & 23.66   & 41  & -92  
\nl

107947 & 207575  & \nodata & \nodata & 21 52 09.7 & -62 03 08
& 7.2  & 0.51    &
F6     & 0.42    & 22.18   & 44  & -94  
\nl

108195 & 207964  & 8352    & \nodata & 21 55 11.3 & -61 53 11
& 5.9  & 0.39    &
F3     & 0.19    & 21.49   & 48  & -93  
\nl

\multicolumn{2}{l}{PPM 366328\tablenotemark{b,c}}
                 & \nodata & ?       & 23 15 01.2 & -63 34 25
& 9.8  & 0.80    &
K0     & 0.15    & 20(guess) & 116 & -44  
\nl

116748 & 222259  & \nodata & $\surd$ & 23 39 39.4 & -69 11 44
& 8.2  & 0.78    &
G5/G8IV& 0.81    & 21.64   & 79  & -60  
\nl

118121 & 224392  & 9062    & $\surd$ & 23 57 35.0 & -64 17 53
& 5.0  & 0.06    &
A1     & \nodata & 20.53   & 83  & -58  
\nl

\cutinhead{1b.  Improbable members}

459    & 67      & \nodata & \nodata & 00 05 28.3 & -61 13 32
& 8.8  & 0.67    &
G5     & \nodata & 18.57   & 87  & -78  
\nl

1399   & \nodata & \nodata & \nodata & 00 17 30.3 & -59 57 04
& 11.3 & 1.4     &
M0     & \nodata & 22.54   & 113 & -40  
\nl

93096  & 175531  & \nodata & \nodata & 18 57 56.6 & -44 58 06
& 9.8  & \nodata &
G8/K0  & \nodata & 15.51   & 27  & -75  
\nl

94051  & 177720  & \nodata & \nodata & 19 08 51.1 & -54 02 17
& 8.7  & 0.56    &
G0     & \nodata & 14.60   & -1  & -66  
\nl

94858  & 180134  & 7297    & \nodata & 19 18 09.8 & -53 23 13
& 6.4  & 0.5     &
F7     & \nodata & 21.94   & 25  & -81  
\nl

94997  & \nodata & \nodata & \nodata & 19 19 49.6 & -53 43 13
& 12.1 & 1.58    &
M3     & \nodata & 16.67   & 14  & -90  
\nl

95302  & 181516  & \nodata & \nodata & 19 23 20.5 & -50 41 20
& 9.0  & 0.74    &
G6IV   & \nodata & 13.24   & 31  & -87  
\nl

97705  & 187101  & \nodata & \nodata & 19 51 23.6 & -58 30 34
& 8.0  & 0.58    &
F8/G0  & \nodata & 14.73   & 36  & -85  
\nl

101636 & 195818  & \nodata & \nodata & 20 36 02.3 & -54 56 28
& 8.6  & 0.58    &
G0     & \nodata & 15.16   & 50  & -84  
\nl

101844 & \nodata & \nodata & \nodata & 20 38 19.4 & -55 36 19
& 11.36& 1.42    &
K4     & \nodata & 31.24   & 14  & -79  
\nl

103438 & 199065  & \nodata & \nodata & 20 57 22.4 & -59 04 33
& 7.95 & 0.66    &
G2/G5  & 0.052   & 19.63   & 12  & -59  
\nl

104256 & 200676  & \nodata & \nodata & 21 07 17.5 & -57 01 55
& 8.8  & 0.82    &
K1     & \nodata & 18.69   & 35  & -112 
\nl

107806 & 207377  & \nodata & \nodata & 21 50 23.7 & -58 18 17
& 7.9  & 0.73    &
G6     & \nodata & 24.46   & 50  & -93  
\nl

109612 & 210507  & \nodata & \nodata & 22 12 16.8 & -54 58 40
& 9.66 & 0.95    &
K3     & \nodata & 20.39   & 108 & -67  
\nl

114236 & 218340  & \nodata & \nodata & 23 08 12.2 & -63 37 41
& 8.4  & 0.62    &
G3     & \nodata & 17.61   & 101 & -63  
\nl

\enddata

\tablecomments{Position, apparent V magnitudes (m$_v$) color (B-V) and
parallax ($\pi$ in milliarcsec) are from the Hipparcos catalog unless
otherwise noted.  Spectral types (SpT) are taken from various sources
in the literature, luminosity classes are main sequence (V) unless
otherwise noted.  A check in the ``nuclear member'' column indicates
the star is likely a member of the Tucanae Association (or nucleus), a
blank indicates possible membership in the ``stream''.  The ROSAT
fluxes in counts per second (cps) are from the ROSAT Bright Source
Catalog (Voges et al. 1998) and the X-ray positions match the stellar
positions to within 20$\arcsec$, unless otherwise noted.  Proper
motions are an average of the Hipparcos and PPM values if both exist.}

\tablenotetext{a}{Star has definite or possible far-infrared excesses
as measured by IRAS.}

\tablenotetext{b}{Possible member.  Space motion is somewhat
descrepant or star doesn't show expected signs of youth.}

\tablenotetext{c}{PPM 366328 is included as a possible member despite
lack of a Hipparcos parallax because its high X-ray
flux indicates potential youth and its photometric distance is
$\sim$50 pc.  Proper motion and m$_v$ are from PPM catalog.}



   
\end{deluxetable}

\begin{deluxetable}{lccccccc}

\tablefontsize{\scriptsize}

\tablenum{2}
\tablecolumns{8}  

\tableheadfrac{0.01}

\tablecaption{Measured and derived quantities \label{Tab2}}
\tablehead{
    \colhead{}    &
    \colhead{H$\alpha$}  &
    \colhead{Li 6708\AA}    &
    \colhead{{\it v sin i}}    &
    \colhead{RV}    &
    \multicolumn{3}{c}{Space Mot. (km s$^{-1}$)}  
\\
    \colhead{HIP}   &
    \colhead{profile}  &
    \colhead{EW m\AA}   &
    \colhead{km s$^{-1}$}    &
    \colhead{km s$^{-1}$}    &
    \colhead{U}  &
    \colhead{V}  & 
    \colhead{W}   
}
    
\startdata

\tablevspace{-0.1in}
\cutinhead{2a. Probable and possible members of the nucleus or stream}

1481    & filled? & 125     & 18.4    &
+7.0    & -8.9    & -19.8   & -1.5    \nl

1910    & EW = -2.2 & 210   & 18      &
+4.0    & -12.6   & -19.4   & +0.3     \nl

1993    & EW = -1.2\tablenotemark{a}   
                  & $<$50   & 17      &
+7.0    & -6.5    & -19.0   &  -1.0   \nl

2484    & \nodata & \nodata & 107     &
+9.0(0,10,14)\tablenotemark{b} 
        & -8.3   & -22.4    & -1.7    \nl



2487\tablenotemark{c}
        & \nodata & 18      & 6.1\tablenotemark{c} &
+9.0(-10.5\&+18,8,13)   
        & -13.0   & -25.9   &  -1.7    \nl



2578    & \nodata & \nodata & 78      &
+7.0 (1,5,7.5,14)
        & -11.0 & -22.0     & -0.6    \nl



2729    & EW = -2.0 & 350   & 110     &
-1.0\tablenotemark{b}
        & -11.5   & -18.6   & +6.9 \nl

92680   & filled   & 260    & 63       &
+0.0 (0,-14,0,4,-3)\tablenotemark{b}
        & -7.7    & -16.4   & -9.4 \nl



93815\tablenotemark{d}      
        & filled  & $\lesssim$70\tablenotemark{e}  
                            & 26.3\tablenotemark{e} &
+3 (+90\&-82,+2)\tablenotemark{b,d}       
        & -9.7    & -24.7   & -13.9   \nl

\nodata & filled  & $<$25\tablenotemark{f}  
                            & 36.6\tablenotemark{f} &
\nodata & \nodata & \nodata & \nodata \nl

        
95261   & \nodata & \nodata & very large\tablenotemark{g}&
-2.0 (-17,13)\tablenotemark{b}
        & -10.9   & -14.6   & -7.9    \nl


95270   & filled? & 125     & 15.7    &
-0.5    & -10.1   & -15.7   & -9.1      \nl

99803-SW
        & filled   & $<$30  & 33      &
-18.0   & -28.7   & -20.2   & +1.8     \nl

99803-NE
        & filled  & $<$45   & 30      &
-16.0   & \nodata & \nodata & \nodata \nl

100751  & \nodata & \nodata & 16      &
+2.0 (3,2)\tablenotemark{b} 
        & -6.5    & -22.6   & -1.7 \nl


104308  & \nodata & \nodata & $>$100(?) &
-10.0\tablenotemark{b}
        & -17.6   & -22.9   & +3.6     \nl

105388  & filled  & 205     & 13.3    &
-1.0 (-1,-5)   
        & -7.9    & -20.0   & -0.8 \nl


105404  & filled  & 150     & 12.8     &
+6.0    & -3.8    & -23.8   & -6.0    \nl

107345  & EW = -1.3\tablenotemark{a}
                  & $<$40   & 14       &
+2.0    & -7.8    & -18.7 & -1.0 \nl

107947  & filled  & 110     & 30       &
+3.0    & -8.3    & -20.8   & -1.2    \nl

108195  & \nodata & 100     & 110      &
-3.0 (-7,1)\tablenotemark{b} 
        & -12.8   & -19.3   & +2.6     \nl


PPM 366328       
        & filled\tablenotemark{h} 
                  & ?       & very large? &
-5.0\tablenotemark{b}
        & -25.1   & -16.0 & -0.7 \nl

116748-S
        & filled  & 215     & 15.7    &
+7.5    & -9.7    & -20.7   & -1.9    \nl

116748-N
        & filled  & 220     & 13.4    &
+6.0    & -9.5    & -21.9   & -1.0 \nl


118121  & \nodata & \nodata & 152\tablenotemark{g} &
+0.6 (-3)\tablenotemark{j}
        & -13.1   & -19.1   & +3.4     \nl


\tablevspace{-0.05in}

\cutinhead{2b.  Improbable members}

459     & normal  & 12      & 5.1     & 
+7.0    & -10.7   & -28.7   & +1.2     \nl

1399    & normal  & $<$30   & $\sim$ 7 &
-4.5    & -19.3   & -16.1   & +5.1 \nl

93096   & normal  & $<$25   & 5.9     &
-13.5   & -20.5   & -16.1   & -9.9 \nl

94051   & normal  & 37      & 5       &
-30.0   & -34.2   & -11.3   & +7.8 \nl

94858   & normal  & $<$10   & 5.9     &
-22.5 (-22.5,-24)
        & -27.7   & -8.6    & +1.4 \nl


94997   & normal  & $<$30   & $\sim$ 13 &
+18.0   & +4.9    & -26.7   & -16.1   \nl

95302   & normal  & $<$15   & 5.5     &
+32.5   & +14.8   & -32.1   & -30.0   \nl

97705   & normal  & 62      & 5.5     &
+17.5   & -1.8    & -27.9   & -20.2  \nl

101636  & normal  & 65      & 4.4     &
-22.5   & -33.1   & -18.4   & +2.4     \nl

101844  & normal? & $<$40   & $\sim$ 8 &
-26.0   & -24.1   & -5.0    & +14.8    \nl

103438  & normal  & 65      & 5.5     &
+11.0   & +2.3    & -16.7   & -6.9    \nl

104256  & normal  & $<$20   & 7.0     &
+22.0   & +3.7    & -33.0   & -16.3   \nl

107806  & normal  & $<$20   & 9.2     &
+13.5   & -0.8    & -22.1   & -10.6   \nl

109612  & normal? & $<$15   & 6.5     &
-10.5   & -26.6   & -16.6   & -1.7    \nl

114236  & normal  & 23      & 4.9     &
+4.0    & -20.1   & -24.8   & -4.9    \nl

\tablevspace{-0.05in}

\cutinhead{Nearby moving groups}

TWA      & \nodata & 
\nodata & \nodata & \nodata & -11 & -18 & -5 \nl

Local Assoc. & \nodata  & 
\nodata & \nodata & \nodata & -11 & -21 & -11 \nl

UMa      & \nodata &
\nodata & \nodata & \nodata & +13 & +1 & -8 \nl

Hyades   & \nodata & 
\nodata & \nodata & \nodata & -40 & -16 & -3 \nl

\enddata

\tablecomments{In the H$\alpha$ column, ``Normal'' indicates an
absorption line similar to inactive stars of comparable spectral type
(e.g. see figures in Soderblom et al. 1998).  ``Filled'' indicates a
shallow absorption or completely absent H$\alpha$ feature.  Negative
equivalent widths (EW) indicate an emission line.  Stars without
entries in the H$\alpha$, Li and FWHM columns are A- and B-type stars
with broad H$\alpha$ lines and few other photospheric features.  The
Li equivalent widths (EW) are not corrected for possible contamination
from FeI 6707.44\AA~line whose contribution is typically $\lesssim$ 10
m\AA~for F- and G-type stars and $\lesssim$ 25m\AA~for K- and M-type
stars.  First listed radial velocity (RV) is used to calculate UVW.
This value is generally our CTIO measurement.  If additional values
appear, the first value in parentheses () is our CTIO measurement
followed by alternative measurements which appear in the literature.
CTIO measured velocities have an error of less than 2 km/s unless
otherwise noted.  Heliocentric UVW space motions have been computed
according to the equations of Johnson and Soderblom (1987), positive U
in the direction of the galactic center, V the direction of galactic
rotation and W toward the North Galactic Pole.  The space motions for
moving groups are from Jefferies 1995 (Local Assoc. taken from Eggen
1992), Soderblom \& Mayor 1993 (UMa \& Hyades) and Webb, Reid \&
Zuckerman 2000 (TW Hya Assoc.= TWA) and are included for comparison.}

\tablenotetext{a}{double peaked H$\alpha$ profile}

\tablenotetext{b}{Error of the RV measurement is significantly larger
than typical because the star is a rapid rotator and/or an early
spectral type with few spectral features.  Uncertainty of these
measurements is typically 10 km/s.}  

\tablenotetext{c}{The BSC lists HIP 2487 as a 0\farcs 4, A2 + A7
binary system, several ionized metal lines are double peaked and the
red peak is not well fit by a gaussian profile.  Nonetheless we
measure velocities of the components at -10.5 km/s and +18 km/s (with
a large uncertainty).  For calculation of UVW, we adopt an average of
our weighted mean system velocity and previously measured velocities.
{\it v sin i} = 6.1 applies to the blue shifted star.  The red shifted
component, based on deblending of only four lines, appears to
be somewhat more rapidly rotating but the non-gaussian profile
suggests the situation is more complicated.  All measurements as of
HJD 2451414.934.}

\tablenotetext{d}{Double-lined spectroscopic binary.  Strength of
photospheric features is approximately equal and components are at RV
= -82 and +90 km/s on HJD = 2451415.472. If stars are equal mass,
average system velocity = +4 km/s, also previously measured to be +2
km/s.}

\tablenotetext{e}{blue shifted component}

\tablenotetext{f}{red shifted component}


\tablenotetext{g}{Very broad H$\alpha$ profile}

\tablenotetext{h}{Extreme rapid rotator?  Spectrum is essentially
featureless.  Broad shallow depressions appear at the positions of a
couple of strong spectral lines which are likely rotationally
broadened features.}

\tablenotetext{j}{We adopt RV = 0.6 km s$^{-1}$ from Grenier et al
(1999) in preference to our measurement of -3 km s$^{-1}$ which has a
larger uncertainty}

\end{deluxetable}

    Properties of the 37 star systems we observed at CTIO are listed
in Tables 1 and 2.  The calculated UVW and the age indicators usually
agree in the sense that the stars with space motions similar to those
of stars in the youthful TWA (Soderblom, King \& Henry 1998; Webb et
al 2000), Beta Pic moving group (Barrado y Navascues et al 1999), and
Local Association (Jeffries 1995), usually also have additional
indications of youth.  Youth is deduced from one or more of the
following characteristics: ROSAT All-Sky Survey Bright Source Catalog
(RASS-BSC, Voges et al. 1998) X-ray source, strong lithium
6708\AA~absorption, H$\alpha$ emission or weak (filled-in) absorption,
rapid rotation, IRAS far-infrared excess emission, and, for the A- and
late-B type stars, location near the bottom envelope of brightness of
stars of comparable spectral type i.e. on or near the zero-age main
sequence, (Jura et al 1998; Lowrance et al 2000).  

    We interprete Tables 1a and 2a in the following way.  Nine or ten
star systems near zero hours right ascension are likely to be part of
a small stellar association similar to the TWA or the U Ma nucleus
(see Figure 1c).  These 10 stars are indicated in column 4 of Table 1a
and we dub them "the Tucanae Association"; their distribution is shown
in Figure 1a.  Other stars in the Tables and Figures, may be placed
into one of the following catagories: (1) a member of the stream of
nearby, young Local Association stars, or (2) a star with no obvious
indications of youth that happens to have distance from Earth and
proper motions similar to those of the Tucanae Association.  Some of
these stars may indeed be Tucanae Association members without
signatures of youth (e.g. see discussion of HD 207129 below).  An
excess of stars with similar proper motions and distance from Earth
seems to exist in the Tucanae region, indicating some stars listed in
Table 1b may be Tucanae members.  But individually, each of these
stars is unlikely to be a member.

\section{Results \& Discussion}

    Ages have been deduced for a few stars in Table 1a.  For example,
HIP 92680 (PZ Tel) is above the main sequence and estimated to be
15-20 million years old (Favata et al 1998; Soderblom et al. 1998).
With the NICMOS camera on HST, Lowrance et al. (2000) discovered a
late M-type object within 4$\arcsec$ of the A-type star HR 7329; if a
companion, which appears very likely, then the object is a brown dwarf
of age $\sim$ 20 Myr.  Other stars with similar space motions and ages
around 20 million years have been identified close to the Sun, for
example, Beta Pictoris (Jura et al 1993; Barrado y Navascues et al
1999) and Gliese 799 and 803 (Barrado y Navascues et al 1999).

    The Tucanae Association itself is probably somewhat older than 20
million years.  We tentatively assign an age of 40 Myrs based on the
strength of H$\alpha$ emission lines seen in HIP 1910, 1993, and 2729
which are brighter than in stars of similar spectral type in the
$\alpha$ Per cluster (see Figure 16 in Prosser 1992) whose age has
recently been estimated to be 90 +/- 10 Myr (Stauffer et al 1999).
Other indicators of age, the presence of rapid rotators, stars with
high Li abundance and or large X-ray flux, all agree with this
estimate but are somewhat less diagnostic.  For example, the X-ray
counts per second of the K- and M-type stars in Table 1a imply
L$_x$/L$_{bol}$ comparable to that of late-type stars in very young
clusters (see Fig. 1 in Kastner et al. 1997).  And L$_x$/L$_{bol}$ of
the late F- and G-type stars in Table 1a are typically three orders of
magnitude larger than L$_x$/L$_{bol}$ of the Sun (Fleming 1999) which
suggests an age younger than the Pleiades (see Fig. 2 in Gaidos 1998).

	Over 40 Myr, dispersion as small as 1 km/s in the critical V
component of space velocity would lead to a 40 pc separation between
stars.  Since no such separation is present among the nuclear members
(Fig. 1a), either the range in V given in Table 2a is due to
measurement error or the stars are younger than 40 Myr or both.
Measurement errors are characterized by, for example, differences
between Hipparcos and PPM proper motions and the large uncertainty in
radial velocity for A- and B-type members.  Also, close companions in
unrecognized binary systems will generate orbital motion that could
shift the measured value of V away from the true V velocity of the
binary system.

    We chose the stars in Table 1 for further study at CTIO because of
similar distances from Earth and proper motions.  However, once radial
velocities were measured it became clear that many of these stars
share similar space motions (UVW) with other very young nearby stars
found in very different directions.  The mean UVW for the nine likely
member systems of the Tucanae nucleus (PPM 366328 not included) is
(-10.5, -20.8, +0.3) $\pm$ (2.3, 2.4, 3.0).  For comparison, UVW for
the TWA is -11, -18, -5 (Webb et al 2000) and, for the Beta Pic moving
group, -10.3, -16.5, -10.2 (Barrado y Navascues et al 1999).  In
addition, we calculate UVW = -11, -18, -10 for $\eta$ Cha, the
brightest member of the recently identified, young, compact $\eta$
Chamaeleontis cluster (Mamajeck, Lawson \& Feigelson 1999).  Similar
space motions are also evident for some stars with far-IR excess
emission as measured by IRAS or having strong lithium 6708\AA~lines
(Jeffries 1995).  As noted by Jeffries, many of these lithium stars
have space motions similar to that of Eggen's Local Association (U,V,W
= -11, -21, -11).  Four such stars, HD 172555, HD 195627, HD 207129
and HD 10647, are plotted as asterisks on Figures 1a and 1b.  These
four stars were not targeted by us for CTIO observations because they
are significantly closer to Earth than stars in Table 1 and we did not
initially recognize them as potential members of a common stream.

    The G-type star HD 207129 is of special interest because it is
surrounded by a cold dust ring detected by IRAS and is only 15.6 pc
from Earth.  Jourdain de Muizon et al (1999) argue that this star is
4.7 Gyrs old and construct a corresponding model for evolution of the
dust ring. In contrast, we believe that HD 207129, UVW = -13.7, -22.3,
+0.6, is actually a member of the Tucanae stream and probably only
about as old as the Tucanae Association.  Stars with space motions
within the range encompassing the young groups listed above, (-15,
-23, -13) $<$ (U, V, W) $<$ (-9, -16, +3), comprise less than 2\% of
the stars in Gleise's Catalog of Nearby Stars.  Thus the chance that
HD 207129 is as old as 4.7 Gyrs and yet have a space motion so similar
to many very young stars is small.  Also, location of HD 207129 in the
same direction as the Tucanae stream stars seen in Figure 1 supports
the idea that they are kinematically associated.  This strikes us as
more compelling evidence for youth than the weak Ca II K-line
emission, relied on by Jourdain de Muizon, as an indicator of a much
older star.  Not all young stars have activity in the Ca II lines.
For example, the very young, F-type star HD 135344 which has a huge
far-IR excess and associated CO rotational emission (Zuckerman,
Forveille, \& Kastner 1995), has no Ca activity (Duncan, Barlow \&
Ryan 1997).  Finally, we note that the intrinsic X-ray luminosity of
HD 207129 as measured by ROSAT is about 10 times that of the Sun.

    Many papers published recently describe field stars with high
lithium abundance, large X-ray fluxes, and other indicators suggestive
of youth, but no consistent or compelling picture has been established
for the solar vicinity.  We believe it is now possible to paint a
plausible picture of the recent star formation history of the present
solar neighborhood.

    Between 10 and 40 million years ago in a co-moving frame centered
near the present position of the Sun, an ensemble of molecular clouds
were forming stars at a modest rate.  The spectral types of these
stars ranged primarily from A to M, but included a few B-type stars
also.  About 10 Myrs ago, the most massive of the B-type stars
exploded as a supernova at about the time that stars in the TWA were
forming.  This event terminated the star formation episodes and helped
to generate a very low density region with radius of order 70 pc in
most directions from the present position of the Sun (Welsh, Crifo \&
Lallemont 1998 and references therein). Thus we now have a ``150 pc
conspiracy'' whereby molecular clouds (Taurus, Lupus, Cha, Sco, Oph)
are seen in various directions, typically $\sim$150 from Earth, and,
like the star-forming clouds 10-40 million years ago, mostly at
negative declinations.  If the rate of supernovae in the Galaxy is one
per 50 years, then in a typical sphere of radius 70 pc, a supernova
will explode every $\sim$10$^7$ years.  Ten million years ago, the Sun
would have been further than 100 pc from the supernova explosion.

The above picture is consistent with one painted by Elmegreen (1992 \&
1993) which was based on more general considerations pertaining to local
galactic structure and Gould's Belt.  In particular, Elmegreen remarks
that "The local star formation activity began 60 million years ago when
the Carina arm passed through the local gas...This scenario is largely
speculative...".  The recent discoveries of very nearby young star
clusters and field stars in the southern hemisphere, greatly enhances the
likelihood of Elmegreen's speculative scenario.

\section{Conclusions}

    We have found a previously unrecogized southern association -
``the Tucanae Association'' - which is only $\sim$45 pc from Earth,
but not quite as young as the recently established TW Hydrae
Association.  Thus, in the past year, the number of known stellar
associations within 60 pc of Earth has increased from two (the Hyades
and U Ma) to four.  The existence of the Tucanae and TW Hydrae
Associations resolves the mystery of how the Beta Pictoris moving
group can be so young, 20 $\pm$ 10 Myrs (Barrado y Navascues et al
1999) and yet so near to Earth (within 20 pc).  That is, 10-40
million years ago, the region through which the Sun is now passing
experienced a significant era of star formation which produced the
Beta Pictoris group, the two southern associations, and related stream
stars.

    We thank the crew at CTIO including N. Suntzeff, T. Ingerson,
M. Fernandez, A. Gomez, E. Cosgrove \& R. Venegas for critical help
acquiring these data.  A special thanks to M. Smith for his
flexibility with telescope maintenence schedules.  We are grateful to
M.  Jura for calling the Mannings and Barlow paper to our attention
and for pointing out the likelihood of proper motion companions to
some of the stars in Table 1.  We thank Dr. Jura and E. E. Becklin and
P. Lowrance for comments on drafts of the paper. This research was
supported by NSF Grant 9417156 to UCLA and by the UCLA Center for
Astrobiology.

\end{document}